\documentclass[12pt]{article}
\usepackage{graphicx}
\begin{document}

\centerline{\bf Discretized opinion dynamics of Deffuant model on scale-free networks}

\bigskip
\centerline{D. Stauffer$^1$,  A.O. Sousa$^2$ and C. Schulze$^1$}

\centerline{$^1$ Institute for Theoretical Physics, Cologne University}

\centerline{D-50923 K\"oln, Euroland}

\bigskip
\centerline{$^2$Institute for Computer Applications 1 (ICA1), University of Stuttgart}

\centerline{Pfaffenwaldring 27, D-70569 Stuttgart, Euroland}

\bigskip
\centerline{e-mails: stauffer@thp.uni-koeln.de, sousa@ica1.uni-stuttgart.de}

\begin{abstract}
The consensus model of Deffuant et al is simplified by allowing for
many discrete instead of infinitely many continuous opinions, on a directed 
Barab\'asi-Albert network.  A simple scaling law is observed.
We then introduce noise and also use a more realistic network and compare the
results. Finally, we look at a multi-layer model representing various age 
levels, and we include advertising effects. 
\end{abstract}
 

\section{Introduction}
Computer simulation of opinion dynamics (consensus models) (Axelrod 1997;
  Deffuant 2000; Deffuant 2002; Weisbuch 2002; Hegselmann 2002; Hegselmann
  2004; Krause 1997; Sznajd-Weron 2000; Stauffer 2000; Stauffer 2002; Galam
1990; Galam 1997; Stauffer 2003) is an important part of sociophysics (Weidlich 2000;
  Moss de Oliveira 1999; Schweitzer 2003). One checks if, starting from a
  random distribution of opinions (Monte Carlo method), one ends up with a
  consensus or a diversity of final opinions. The simulated people ("agents")
  are located on lattices, on scale-free networks (Albert 2002; Barab\'asi
  2002), or form a purely topological structure where everybody can be
  connected with everybody. For the particular case of the
consensus model of Deffuant et al (Deffuant 2000; Deffuant 2002; Weisbuch
  2002), it was shown that on a Barab\'asi-Albert (BA) network (Stauffer 2004) 
(see also Weisbuch 2004) the number $S$ of different surviving opinions (if no complete
consensus was achieved) was an extensive quantity, i.e. it varied proportional
to the number $N$ of agents, while it is intensive (independent of $N$ for
large $N$) when everybody can be connected to everybody (Ben-Naim 2003). The
literature on Barab\'asi-Albert networks contains many comparisons with reality,
e.g. for the computer networks of the Internet. 

The motivation of the present work is two-fold: We want to have an unambiguous
criterion whether two opinions agree or disagree, and thus use discrete
instead of continuous variables for the opinions, section 2. Then we want
to make the model more realistic by introducing noise representing events
outside the opinion dynamics of Deffuant et al, by using in section 3 
a more realistic network (Davidsen 2002; Holme 2002; Szab\'o 2003) with a
higher clustering coefficient
that the BA network, by allowing for advertising through mass media, and by
taking into account more than one layer in order to implement an age structure;
the last two effects are dealt with in section 4. An appendix gives the 
basic Fortran program.

\begin{figure}[hbt]
\begin{center}
\includegraphics[angle=-90,scale=0.5]{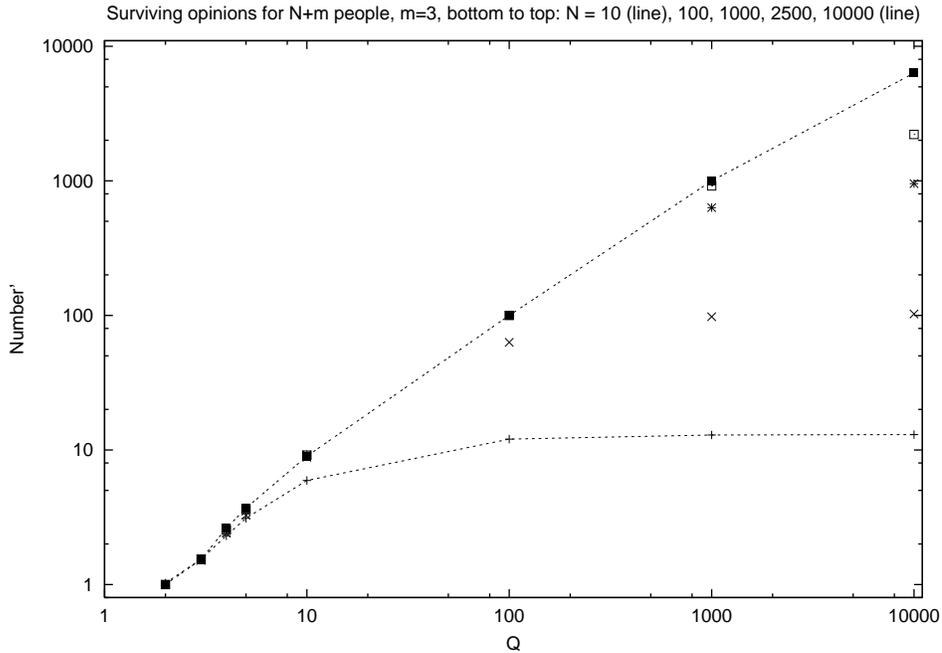}
\end{center}
\caption{Number of different surviving opinions versus total number $Q$ of 
opinions, for various network sizes $N$. Data for $N= 10$ and $N=10^5$ are
connected by lines.
}
\end{figure}

\begin{figure}[hbt]
\begin{center}
\includegraphics[angle=-90,scale=0.5]{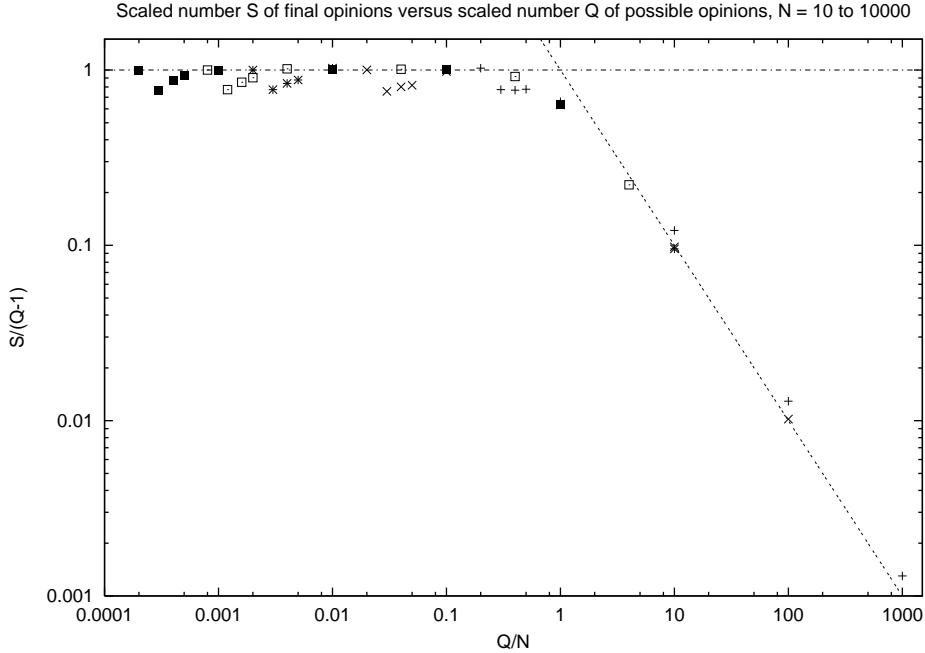}
\end{center}
\caption{Scaled plot of the same data as in Fig.1. The straight lines indicate
the "trivial" scaling limit: Everybody keeps its own opinion in the right
part, and each opinion is shared by many in the left part.
}
\end{figure}

\begin{figure}[hbt]
\begin{center}
\includegraphics[angle=-90,scale=0.5]{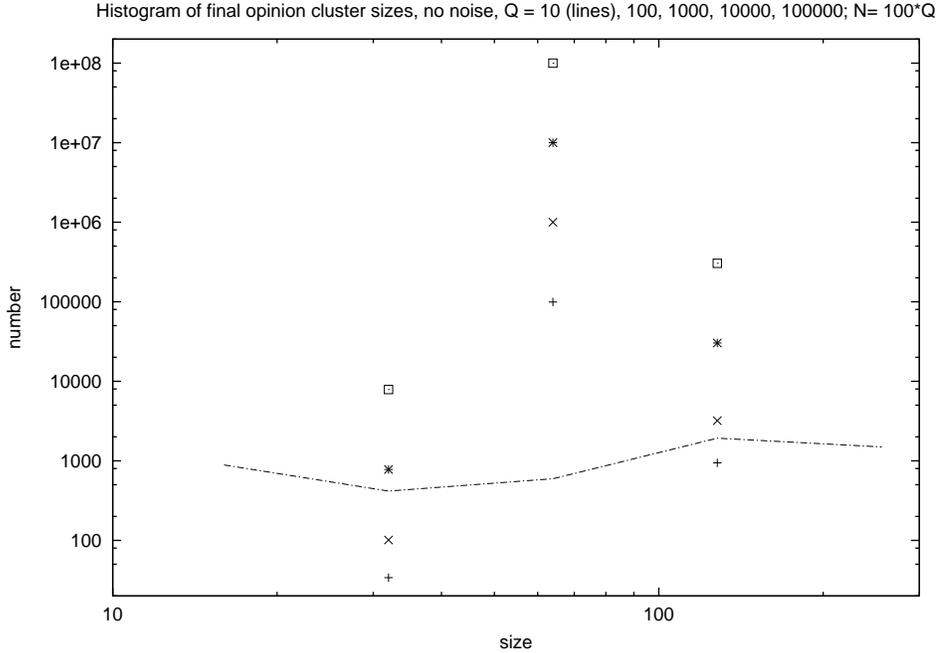}
\end{center}
\caption{
Evidence that cluster numbers for fixed $N/Q$ (here taken as 100) are extensive,
for large systems up to 10 million nodes, with $N$ and $Q$ increasing from bottom
to top. For small systems the results (line) are very different. All data are
summed over 1000 runs.}
\end{figure}

\begin{figure}[hbt]
\begin{center}
\includegraphics[angle=-90,scale=0.5]{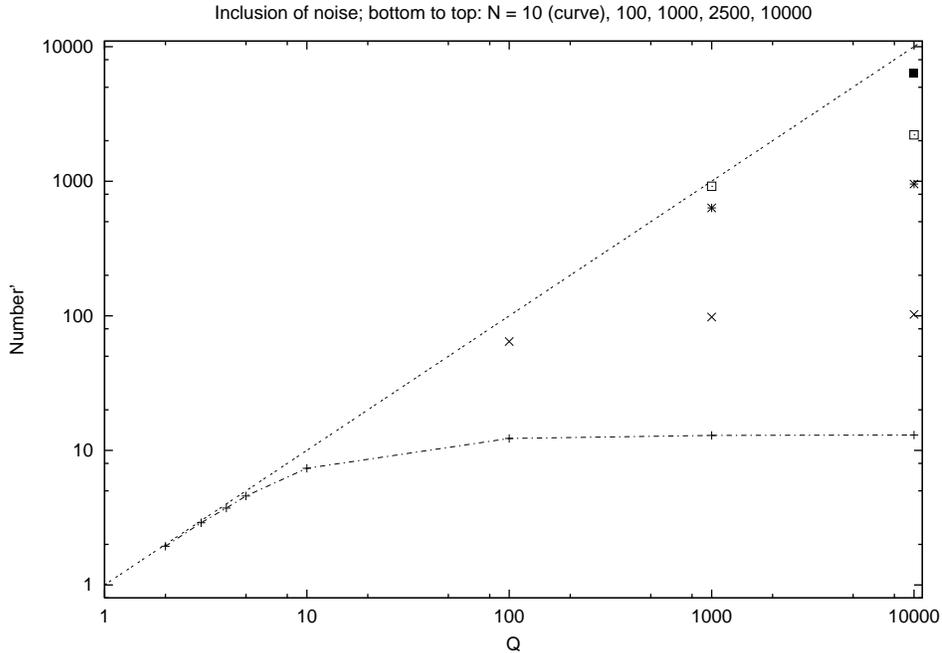}
\end{center}
\caption{As Fig.1, but with noise, for $N = 10$ (curve) to 10000. The straight
line gives $S = Q$. For smaller $Q$ at $N \ge 100 $, no convergence within
$10^6$ iterations was found.}
\end{figure}

\begin{figure}[hbt]
\begin{center}
\includegraphics[angle=-90,scale=0.5]{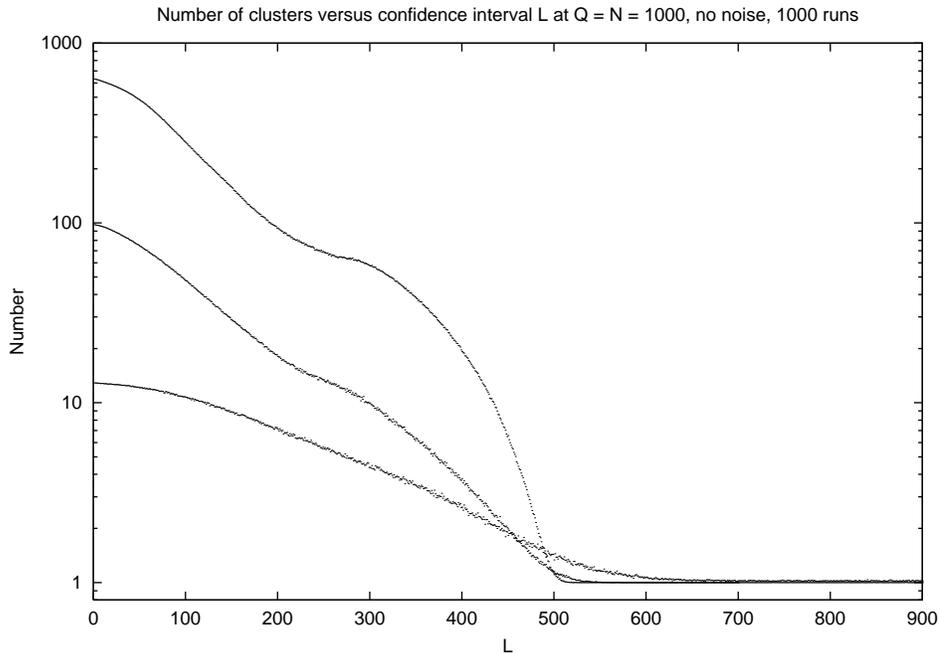}
\end{center}
\caption{As Fig.1, no noise, for $Q=1000$ as a function of the length 
$L = Qd$ of the confidence interval, for $N = 10, 100, 1000$ (from bottom 
to top). The larger $N$ is
the more pronounced is the transition to complete consensus
at  $L = 500$ or $d = 1/2$.
}
\end{figure}

\begin{figure}[hbt]
\begin{center}
\includegraphics[angle=-90,scale=0.33]{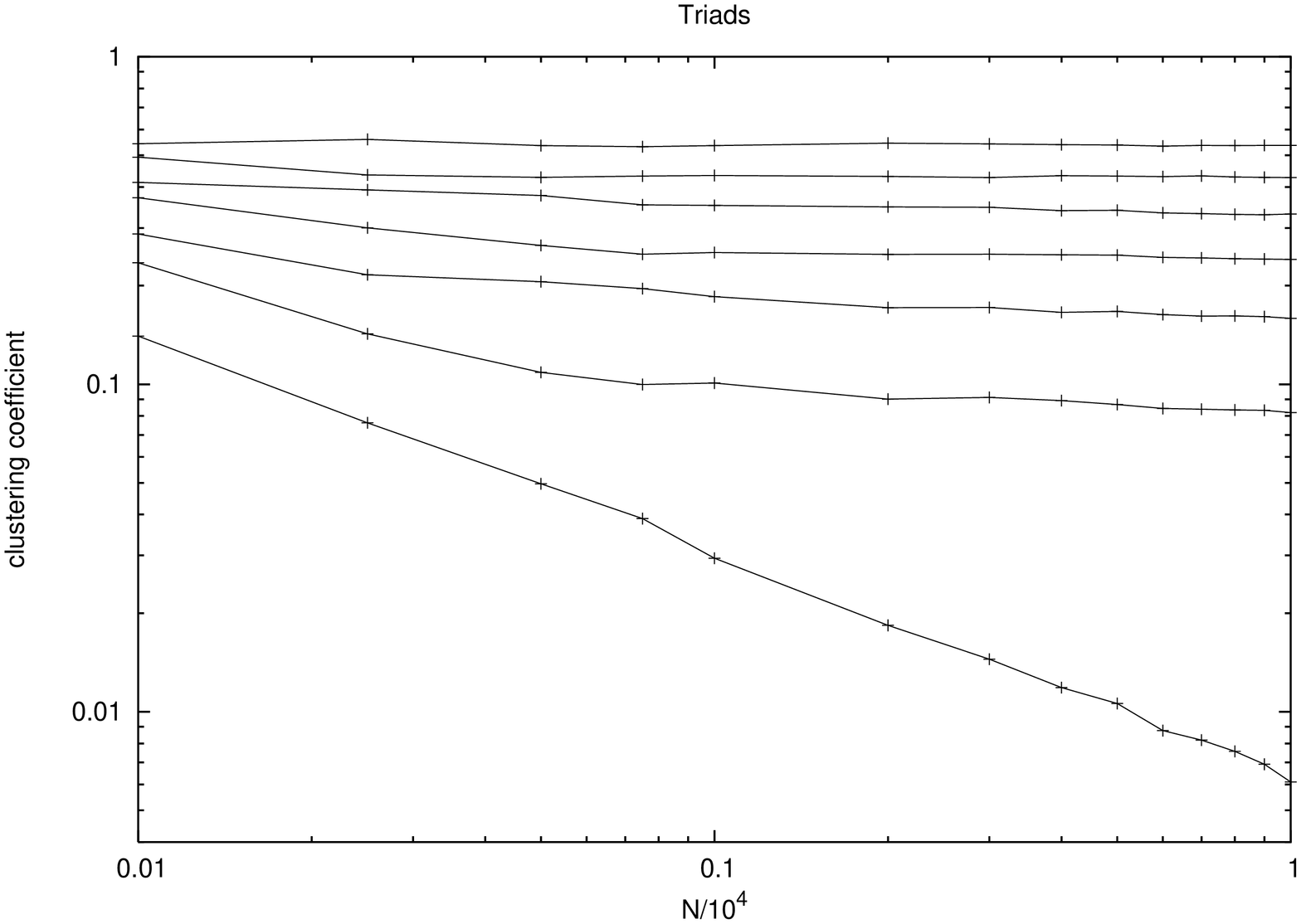}
\includegraphics[angle=-90,scale=0.33]{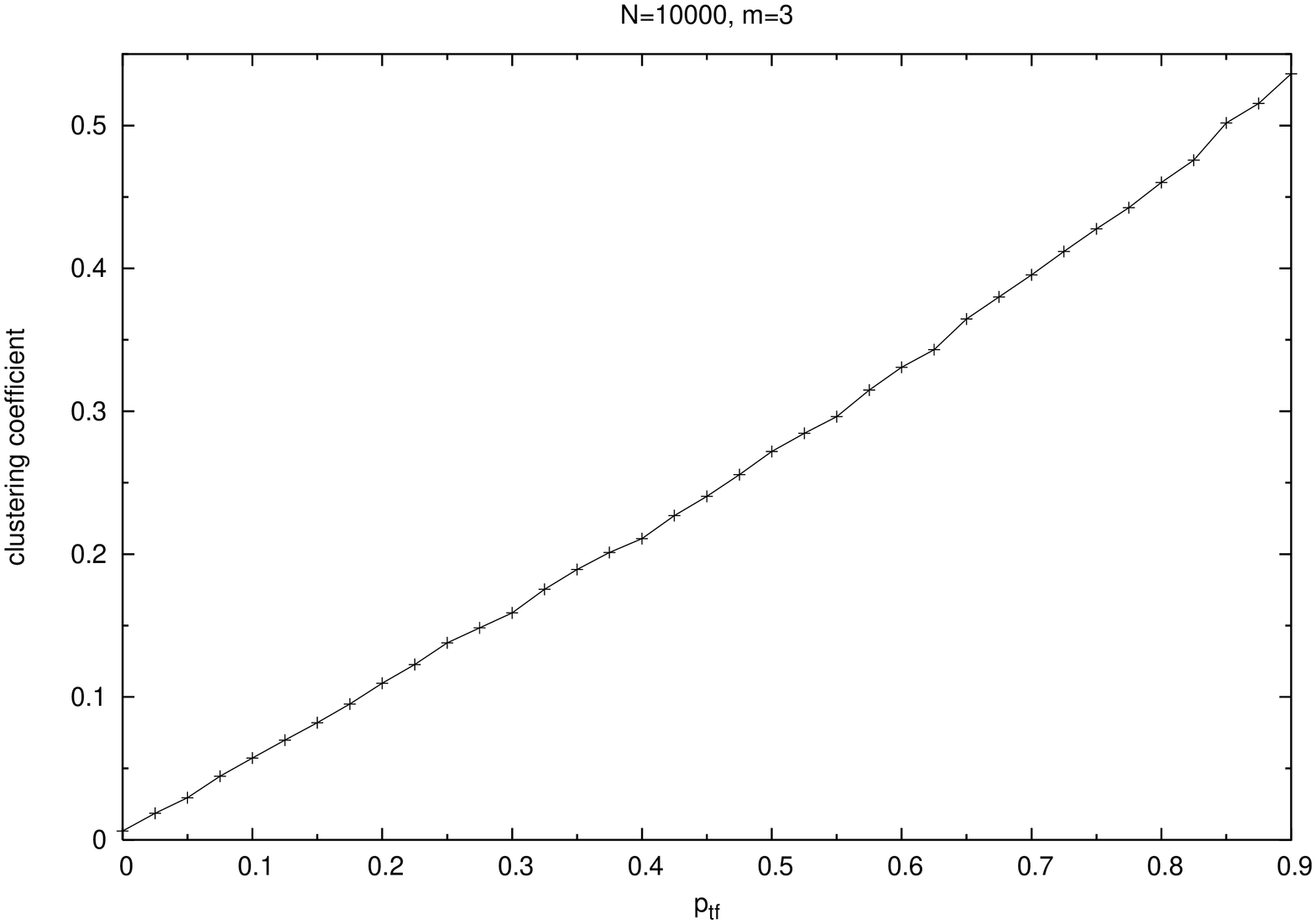}
\includegraphics[angle=-90,scale=0.50]{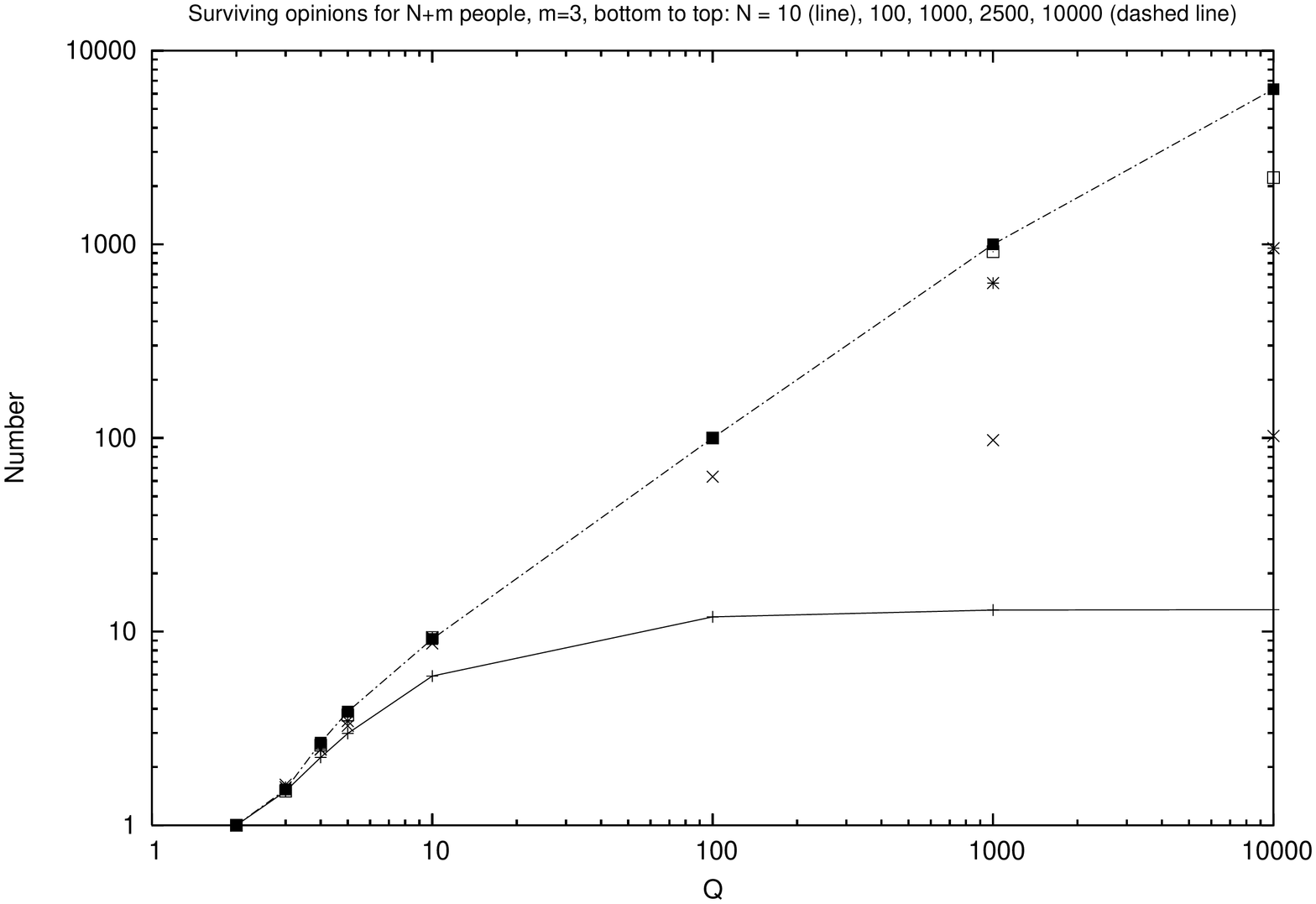}
\end{center}
\caption{Triad formation:
a) Log-log plot: Clustering coefficient versus the network size $N$ at  
probability $p_{\rm tf}$ = 0.00, 0.15, 0.30, 0.45, 0.60, 0.75. 0.90 (from bottom
to top) to perform a triad formation step.
b) Linear plot: Clustering coefficient versus the probability $p_{\rm tf}$ to 
perform a triad formation step at $N = 10^4$.
c) Log-log plot: Number of different surviving opinions versus total number $Q$ 
of opinions, for various network sizes $N$ on a scale-free network with
triad formation step. The triad formation probability is $p_{tf}=0.3$}
\end{figure}

\begin{figure}[hbt]
\begin{center}
\includegraphics[angle=-90,scale=0.5]{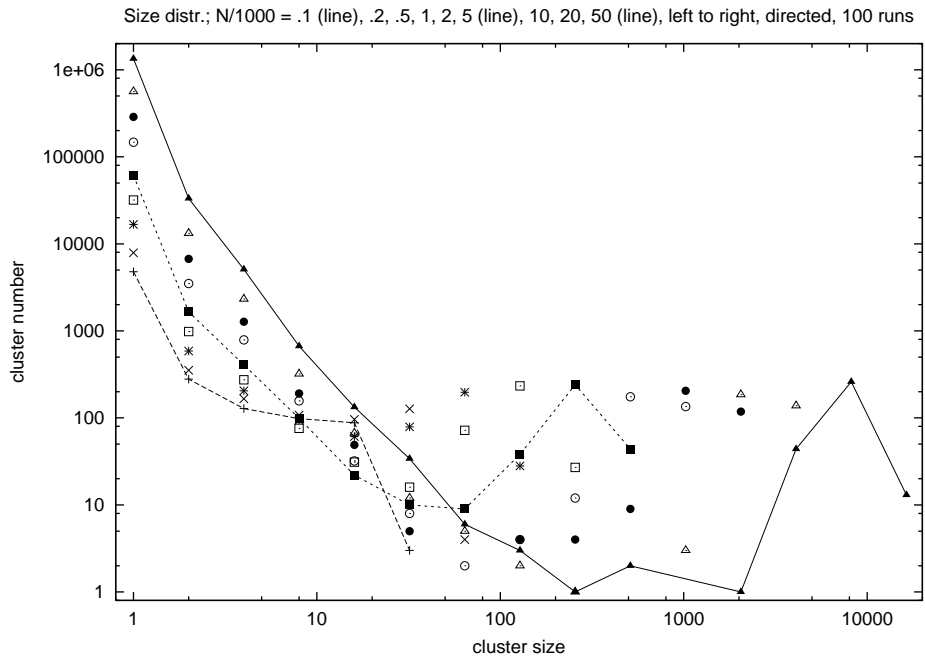}
\end{center}
\caption{Size distribution for the clusters of different surviving opinions,  
summed over 100 samples, for network sizes $N = 1000 \dots 50,000$ on the 
directed a scale-free network with continuous opinions as in the standard
model (Deffuant 2000; Deffuant 2002; Weisbuch 2002).}
\end{figure}

\begin{figure}[hbt]
\begin{center}
\includegraphics[angle=-90,scale=0.43]{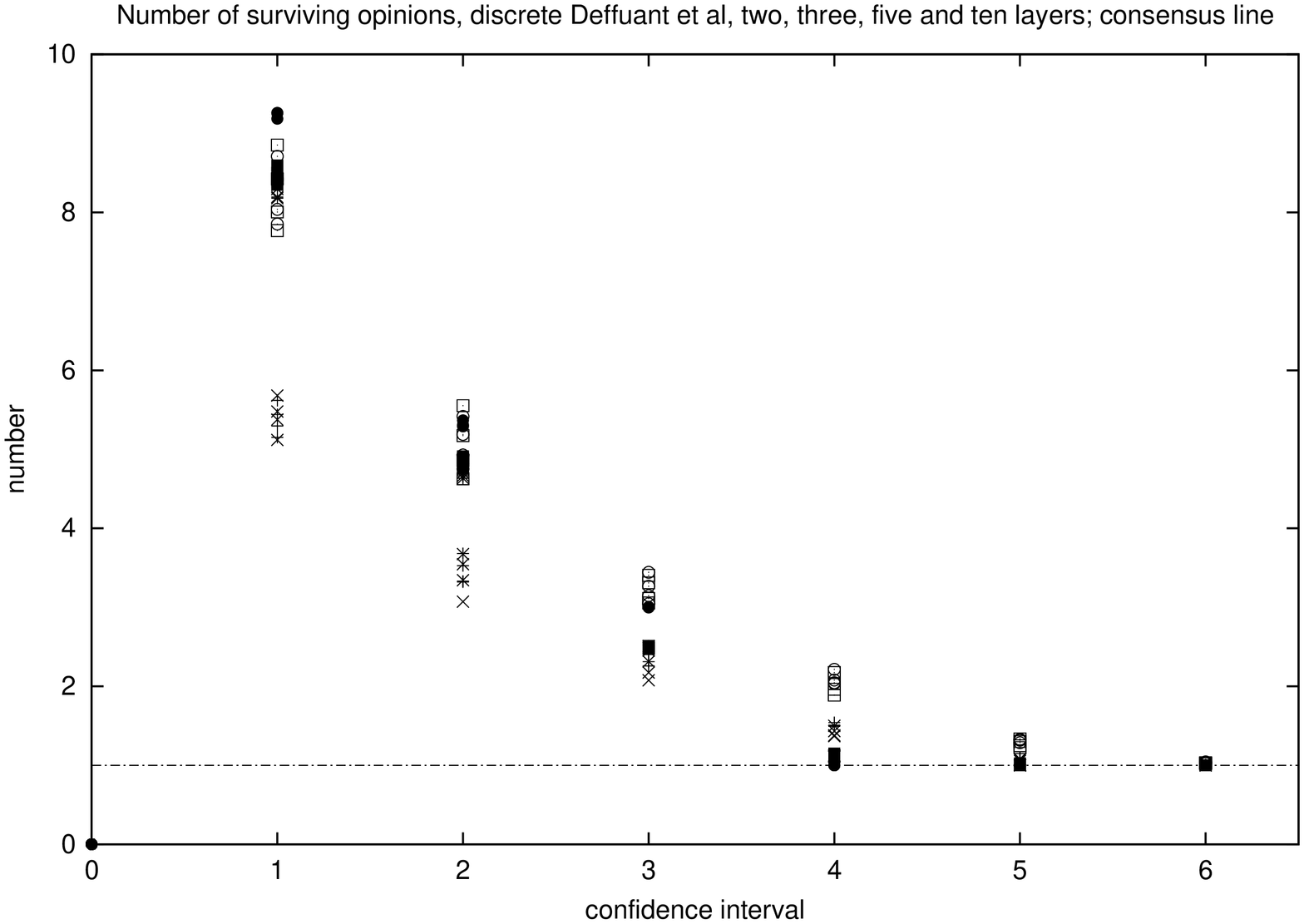}
\includegraphics[angle=-90,scale=0.43]{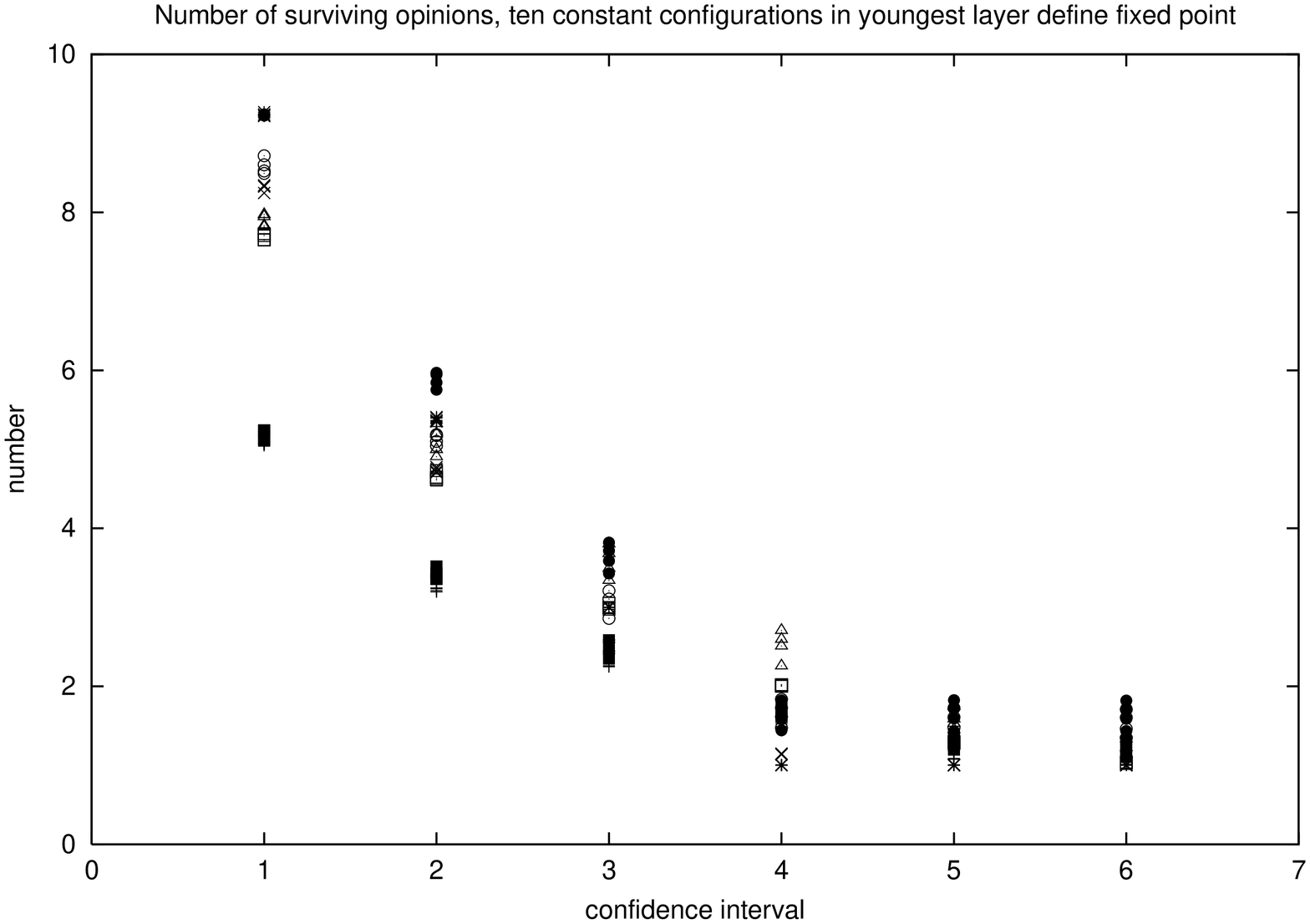}
\end{center}
\caption{Multilayer model:
Number of different surviving opinions 
versus confidence interval $L$ for $Q = 10$ possible opinions. (For $Q = 100$
instead of 10 we plot the number versus $L/10$ instead of versus $L$, also in 
Figs. 9 and 10.) We see complete consensus for $L/Q \ge 0.6$. Part (a) stops
the simulation if in the whole system no one changed opinion during one 
iteration, while in part b we stopped it if the baby layer did not change 
over ten consecutive iterations.}
\end{figure}

\begin{figure}[hbt]
\begin{center}
\includegraphics[angle=-90,scale=0.5]{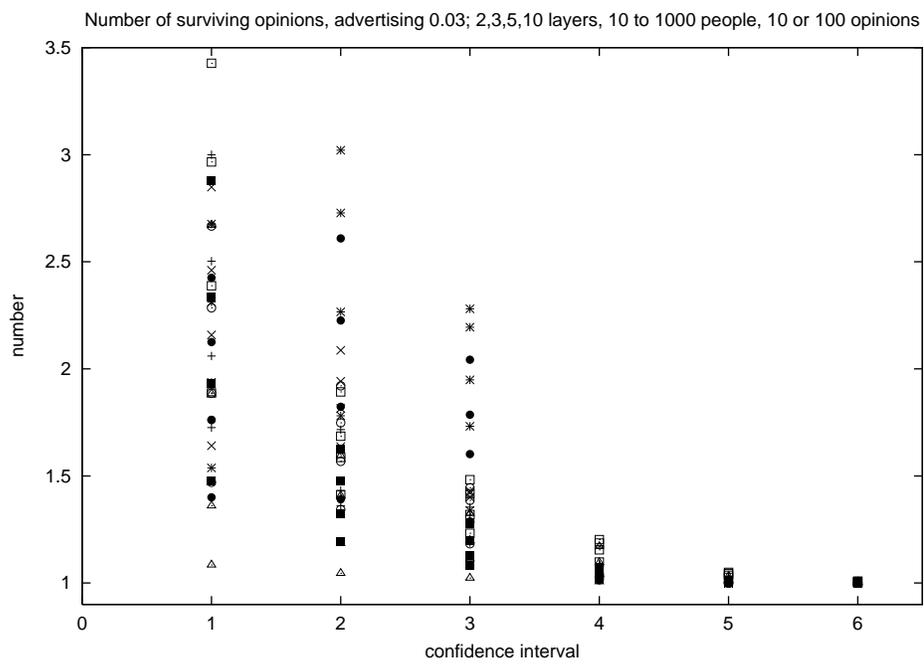}
\end{center}
\caption{Number of different surviving opinions versus $L$ in multilayer model 
in case of advertising. For a monolayer this number increases up to nearly 8.}
\end{figure}

\begin{figure}[hbt]
\begin{center}
\includegraphics[angle=-90,scale=0.5]{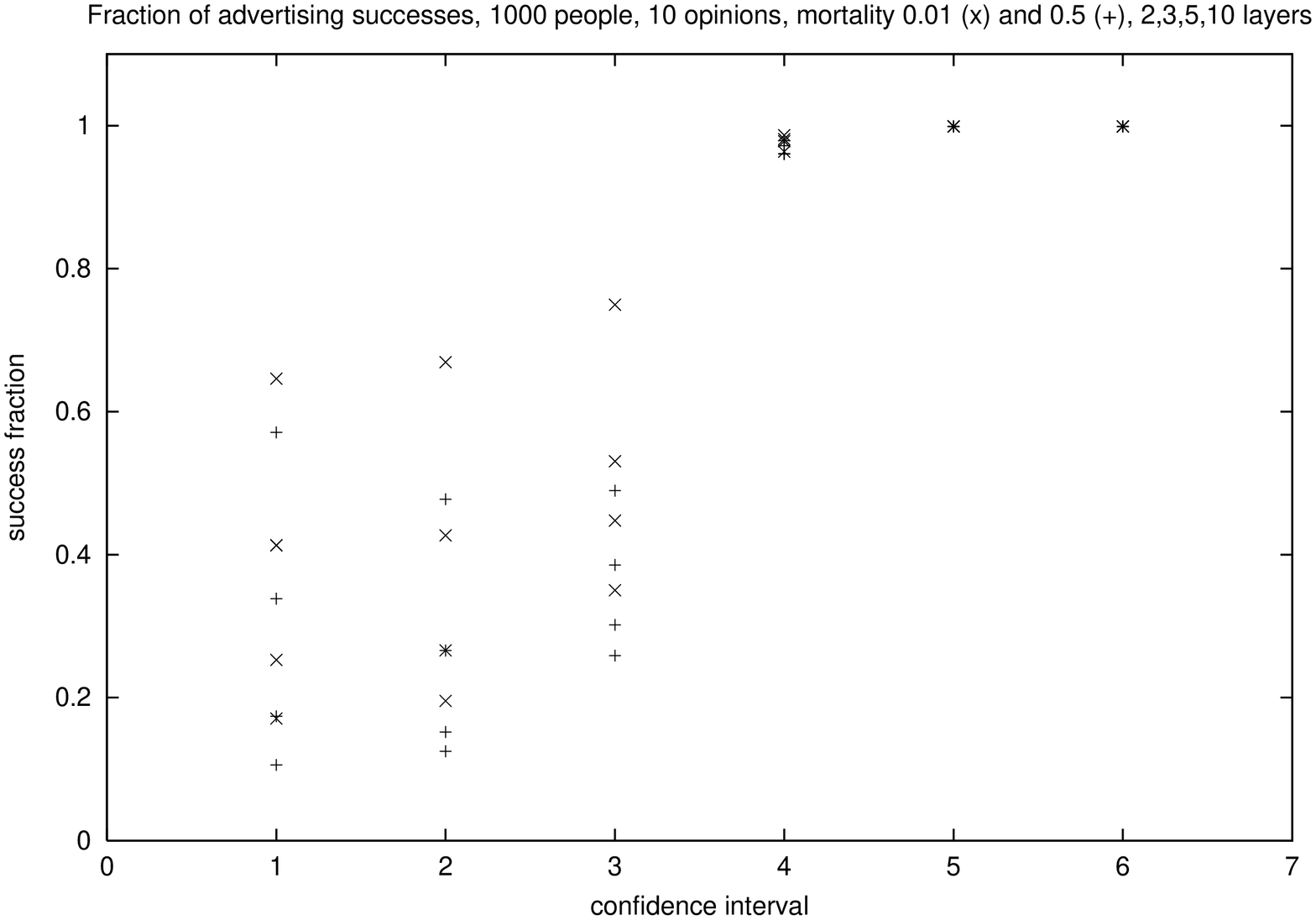}
\end{center}
\caption{Fraction of advertising successes in multilayer model. For a monolayer
the results for $L/Q \ge 0.4$ are similar but for $L/Q = 0.1$ the success
fraction is at most 0.1. And for $Q = 100$ the monolayer success ratio was 
even lower.}
\end{figure}

\section{Basic model and noise}

Instead of allowing for the opinions any real number between 0 and 1, we take
them as discrete numbers $q = 1,2, \dots Q$, as in the Sznajd model 
(Sznajd-Weron 2000; Stauffer 2000; Stauffer 2002). Now it is well defined if two 
opinions differ or agree, while for real numbers it depends on the accuracy of 
the simulation. At first, only people differing by $\pm 1$ in their opinion can 
convince each other (bounded confidence (Hegselmann 2002; Hegselmann 2004; 
Krause 1997, Deffuant 2000; Deffuant 2002; Weisbuch 2002)); thus $1/Q$
corresponds to the confidence interval of the previous models. If two agents 
with opinions differing by one unit talk to each other, randomly one of them
takes the opinion of the other (Axelrod 1997). We put agents on a directed 
Barab\'asi-Albert (Albert 2002; Barab\'asi 2002; Stauffer 2004) network, 
starting with $m=3$ agents connected
with each other and with themselves; thereafter, $N$ agents are added, each of
which selects $m$ pre-existing agents to be connected with. These randomly
selected old agents are {\it not} regarded as connected to the new agent, i.e.
the connections are directed.  In this BA network the number of agents having
$k$ neighbours is known to be proportional to $1/k^3$.  The size of the network
is the number $N$ of agents on it, i.e. the population.

First we construct the network, then we start the opinion dynamics from a random
distribution of opinions. For each iteration we go through all agents in the 
order in which they were added to the network, and each selects randomly one
of the $m$ agents it had chosen before to be connected with. The simulation
stops if no agent changed opinion during one iteration. (About the same results 
are obtained from random instead of regular updating, provided we stop if for
ten consecutive iterations no opinion changed.)

Figure 1 shows that for large $N$ the number $S$ of surviving final opinions
roughly equals $Q$ for not too small $Q$; for $Q=2$, on the other hand, nearly 
always a complete consensus $S=1$ was found. (We averaged over 1000 samples 
except for $N = 10^5$ when only 100 samples were used.) If, however, $Q$ grows
to values closer to $N$, then the finite size of the network is felt:
$S$ is lower than $Q$ and approaches $N+m$, that means everybody keeps its 
own opinion and the simulation stops soon. A finite-size scaling formula
$$S=(Q-1)f(Q/N);\quad f(x \rightarrow 0)=1, \; f(x \rightarrow \infty)=1/x $$
fits reasonably the same data, Fig.2, except for small $Q$. Thus, the 
number $S$ of final opinions is an extensive quantity if $Q$ is varied as $N$,
and it is an intensive quantity if $Q$ is kept constant when 
$N \rightarrow \infty$. In this sense the new results are in between 
the intensive $S$ of (Deffuant 2000; Deffuant 2002; Weisbuch 2002; 
Ben-Naim 2003) and the extensive $S$ of (Stauffer 2004). (Also in contrast
to (Stauffer 2004), the histograms for the number of people sharing the same opinion
show a single peak. As in (Stauffer 2004) people may share the 
same opinion even if they are disconnected.) Fig.3 shows that for a fixed ratio
$N/Q=100$ the cluster numbers are extensive.

Roughly the same results, Fig.4, are obtained if noise is introduced to simulate
outside information in the opinion dynamics. Thus at each iteration, every agent
after the above dynamics shifts the opinion by +1 with probability 1/4, by $-1$ 
with probability 1/4, and keeps it unchanged with probability 1/2. (However,
the opinion cannot leave the interval from 1 to $Q$. The simulation stops if 
without this noise no opinion would have changed.)  Thus the model is robust
against this noise. Of course, now no complete consensus is found, not even at
$Q = 2$. (If we follow (Deffuant 2000; Deffuant 2002; Weisbuch 2002) and 
allow everybody to interact with everybody, the results with noise are nearly 
the same as in Fig.4.)

The present model gets closer to the original Deffuant model if we introduce 
another free parameter $L$ such that two people convince each other if their
opinions do not differ by more than $L$ units; $L=1$ then is our previous
discrete model. If $Q$ and $L$ both go to infinity at constant ratio $d=L/Q$,
then this ratio is the $d$ of Deffuant et al. The parameter $\mu$ of 
(Deffuant 2000; Deffuant 2002; Weisbuch 2002) was taken as $0.1^{1/2}$.  
Fig. 5 shows the variation of the number of surviving opinions with $L$, 
at fixed $Q=1000$ and various $N$.  No noise is used here since noise prevents 
convergence except for very small $L$. For $Q=N=5L$ the number of opinion 
clusters seems to vary proportional to $N$ for large $N$. 

\section{Triads}

Although the Barab\'asi-Albert (Albert 2002; Barab\'asi 2002; Stauffer 2004) 
network has successfully explained the scale-free nature of many networks, 
a striking discrepancy between it and real networks is that the value of the
clustering coefficient - which is the probability that two nearest 
neighbours of the same node are also mutual neighbours - predicted by the
theoretical model decays very fast with the network size and for large 
systems is typically several orders of magnitude lower than found 
empirically (it vanishes in the thermodynamic limit, Fig.6a). In social 
networks (Davidsen 2002; Holme 2002; Szab\'o 2003), for instance, the 
clustering coefficient distribution $C(k)$ exhibits a power-law behaviour, 
$C(k) \propto k^{-\gamma}$, where $k$ is number of neighbours (degree or 
connectivity) of the node and $\gamma \approx 1$ (everyone in the network 
knows each other). 

Very recently, by adding a triad formation step on the
Barab\'asi-Albert prescription, this problem has been surmounted and 
scale-free models with high clustering coefficient have been 
investigated (Davidsen 2002; Holme 2002; Szab\'o 2003). The algorithm 
is defined as follows: First, each new node
performs a preferential attachment step, i.e, it is attached randoml
to one of the existing nodes with a probability proportional to its
degree; then follows a triad formation step with a probability
$p_{tf}$: the new node selects at random a node in the neighbourhood of
the one linked to in the previous preferential attachment step. If all 
neighbours are already connected to it, then a preferential attachment
step is performed (``friends of friends get friends''). In this model,
the original Barab\'asi-Albert network corresponds to the case of 
$p_{tf}=0$. It is expected that a nonzero $P_{tf}$ gives a finite nonzero 
clustering coefficient as $N$ is increased, and with the clustering 
coefficient going to zero when $P_{tf}=0$ (the BA scale-free network model).

Simulations of the discretized version of the Deffuant model on this
network produces similar results (see Fig. 6) to those
obtained using a Barab\'asi-Albert network (Fig. 1). The same behaviour
was found for any value of the probability $p_{\rm {tf}}$ to perform a triad
formation step. Furthermore, the clustering coefficient (Fig. 6b) agrees
with the one  predicted in (Davidsen 2002; Holme 2002; Szab\'o 2003), 
with a nearly linear increase with probability $p_{\rm {tf}}$. 
(Other values of $m$ give qualitatively the same behaviour.)

For completeness we end with some results from the directed model of 
(Stauffer 2004) on the size distribution of opinion clusters, those for the
undirected case are similar. Fig.7 shows one peak for 
small sizes, with cluster numbers increasing proportional to the system
size, and another peak for large sizes of the order of the network size, with 
small size-independent cluster numbers.

\section{Several layers}

Following Schulze (Schulze 2004), we now put $A$ copies of the same directed 
Barab\'asi-Albert network on top of each other, with $N$ agents on every
layer. Each layer corresponds to a certain age cohort, with babies on the
bottom (layer $A$), the oldest old
on the top (layer 1), and intermediate ages in the $A-2$ layers in between.
Each person in every layer dies with probability $p$ at each iteration. In case
of death, the younger people on that position in the network move one layer
up, keeping their opinion, and the lowest layer is occupied by a newly born baby
getting the opinion of the parent. Initially, all people of different ages on
the same position in the network share the same opinion, but later each layer
draws its random numbers independently for random sequential updating. Thus it 
is crucial that the same network appears in $A$ copies on top of each other, 
in order to have a unique identification of ancestors and offspring. We refrain
from comparing the model with university teachers (full professors, associate
professors, ..., down to teaching assistants) waiting for the superior to leave
the job and make it available to younger ones.

Our simulations averaged over 1000 samples and four combinations of the number
$N$ of agents and the number $Q$ of possible opinions: $(N,Q)$ = (10,10), 
(100,10), (1000,10) and (10,1000). We used mortalities $p = 0.01$ and 0.5,
maximum age $A$ = 2, 3, 5, 10, and lengths $L$ of the confidence interval = 1,2,
... 6 for $Q = 10$, and $L = 10, 20, ... 60$ for $Q = 100$. Figure 8 shows 
that only $L/Q$ is really important: If it is 0.6 or higher, all samples
lead to a full consensus with only one opinion surviving in all layers together.
For small $L$ the number $S$ of surviving opinions is seen to be slightly below 
10, the usual maximum number of opinions. Even for $L = 1$ and $Q = 100$ or 1000
(not shown) $S$ 
remains of order ten if $N = 10$; we have to increase $N$ to get larger $S$ 
for larger $Q$. Also with $N = 10000$ and $Q = 10$ the results (not 
shown) look like in Fig.8a. Thus the basic result is similar to the monolayer
results in Sec. 2: 
\clearpage
\noindent
For small $L/Q$ the number $S$ of surviving opinions is 
of the order of $N$ or $Q$, whatever is smaller. The number $A$ of layers
and the mortality $p$ hardly influence the results.

In Fig.8a we let the simulations run, until during one iteration none of the
$A \times N$ agents changed opinion. Figure 8b shows that the results are nearly
the same if we follow (Schulze 2004) and stop the simulation when the baby layer
remained unchanged during ten consecutive iterations; however, for $L/Q \ge 0.6$
instead only one surviving opinion we find on average between one and two.

Finally, we introduce advertising (Schulze 2003; Sznajd-Weron 2003) in 
favour of opinion $q = 1$: 
with 3 percent probability, every agent at every iteration had its opinion 
reduced by one unit. Now $S$ is reduced and depends 
stronger on the various parameters, Fig.9. Again, for a confidence interval 
$L/Q \ge 0.6$ a complete consensus is found. Fig.10 shows for the same runs
the success fraction: With what probability is the final opinion within one 
layer a consensus in the advertised opinion? For $L/Q \ge 0.5$
nearly complete success is seen.
  
\section{Conclusions}

By discretizing the opinions, the simulations of the Deffuant model could be 
simplified and made less ambiguous. Two limits are quite trivial: With
many people and few opinions, nearly all opinions have some followers, and
the number of final opinion clusters nearly agrees with the total number 
of opinions. In the opposite limit of many opinions for few people, nearly
every person forms a separate opinion cluster. For the transition between
these two limits, a simple scaling law is observed for the discretized opinions.
At a fixed ratio of the number of people to the number of opinions, the 
number of final opinion clusters is extensive. Noise and a more realistic 
network with stronger clustering (Davidsen 2002; Holme 2002; Szab\'o 2003)
do not change the results much in the discretized model. An ageing model
with several layers representing different age groups gave results not much
different from those of one single layer, also if advertising is included.

J. Ho\l yst suggested to include noise, and J. Kert\'esz to use the network
of (Davidsen 2002; Holme 2002; Szab\'o 2003). A.O. Sousa thanks a grant from 
Alexander von Humboldt Foundation.
  
\section{Appendix}
To facilitate others to continue this research, to state details unambiguously
and to allow checks for possible errors, the main program used for section 2
is reproduced here. An electronic version is available from 
stauffer@thp.uni-koeln.de as deffuant14.f.

Loop 35 goes over various values of the confidence interval {\tt idis}, 
loop 17 over {\tt nrun} different samples. {\tt ibm} is a random odd integer
with 64 bits. All computations with {\tt derri...} involve the
Derrida-Flyvbjerg (Derrida 1986) parameter omitted from the text for 
simplicity and may be ignored by the reader. (The results for this parameter 
were similar to Weisbuch 2004.)

After initialization, loop 7 connects the initial core of the BA network, and
loop 2 adds to it {\tt max} sites {\tt i} each of which builds {\tt m} directed
bonds to neighbours {\tt neighb(i,new),  new=1,2,...m}, selected randomly 
according to the BA rule with the help of the Kert\'esz {\tt list} of length
{\tt L}. After this network is built up, loop 5 initializes the opinions {\tt 
is} randomly, and loop 9 makes the {\tt maxt} iterations of the Deffuant process. 
In this process for each site {\tt j} in loop 10 randomly a {\tt neighb(j,i)} is 
selected, and if these two agree already or are too far from each other
in their opinion, noting is done: {\tt goto 12}. Otherwise the counter 
{\tt ichange} is increased by one, thus giving the number of opinion pairs 
which have changed during this iteration. If both opinions differ only by
one unit, one of them is selected randomly. If the opinion difference is 
greater than one (but not exceeding the confidence interval {\tt idis}) 
then both opinions change by the same amount {\tt idiff} according to the 
usual Deffuant rule. The lines adding noise to the opinion dynamics are 
commented out in this version. If loop 10 ends going through all sites {\tt j} 
without having changed any opinion, the iterations stop: {\tt goto 11}. 

Now the analysis starts: {\tt nhist(i)} counts how often opinion {\tt i} is
found in the final set of opinions {\tt is}: loops 28 and 29. Loop 27
gives {\tt icount} 
as the number of different surviving opinions, later averaged over many
samples using {\tt ict} to give the average number of different surviving
opinions, i.e. the crucial quantity of this study. {\tt number} is used 
for the binned size distribution {\tt ns}
of opinion clusters, and loops 33, 34, 31 can be ignored as mentioned above.

\begin{verbatim}
      parameter(nsites=10  ,m=3,iseed=4711,maxt=1000000,iq=1000
     1  ,nrun=1000,max=nsites+m, length=1+2*m*nsites+m*m)
      integer*8 ibm,mult
      dimension list(length),neighb(max,m),is(max),nhist(0:iq),
     1 number(max), ns(31),irand(0:3)
      data irand/0,0,-1,1/
      w=sqrt(0.1)
      print 100, iq, nsites, m, iseed, maxt, nrun, w
 100  format('# directed, more confidence', 6i9,f6.3)
      factor=(0.25d0/2147483648.0d0)/2147483648.0d0
      facto2=factor*2*iq
      mult=13**7
      mult=mult*13**6
      ibm=2*iseed-1
      do 35 idis=100,900,100  
      derrisum=0.0
      do 16 i=1,31
 16     ns(i)=0
      ict=0
      do 17 irun=1,nrun
        do 29 n=1,max
 29       number(n)=0
        do 7 i=1,m
          do 7 nn=1,m
            neighb(i,nn)=nn
 7          list((i-1)*m+nn)=nn
        L=m*m
c       All m initial sites are connected with each other and themselves
        do 1 i=m+1,max
          do 2 new=1,m
 4          ibm=ibm*16807 
            j=1+(ibm*factor+0.5)*L
            if(j.le.0.or.j.gt.L) goto 4
            j=list(j)
            list(L+new)=j
            list(L+m+new)=i
 2          neighb(i,new)=j
 1        L=L+2*m
c     end of network and neighbourhood construction, start of opinion change

      n=max
      do 5 i=1,n
        ibm=ibm*16807
 5      is(i)=1+iabs(ibm)*facto2
c     print *, is
      do 9 iter=1,maxt
       ichange=0
       do 10 j=1,n
 6      ibm=ibm*16807
        i=1+(ibm*factor+0.5)*m
        if(i.le.0.or.i.gt.m) goto 6
        i=neighb(j,i)
        if(is(i).eq.is(j) .or. iabs(is(i)-is(j)).gt.idis) goto 12
        ichange=ichange+1
        if(iabs(is(i)-is(j)).eq.1) then
          ibm=ibm*16807
          if(ibm.lt.0) then
            is(i)=is(j)
          else
            is(j)=is(i)
          end if
        else
          idiff=isign(ifix(0.5+w*iabs((is(i)-is(j)))),is(i)-is(j))
          is(j)=is(j)+idiff
          is(i)=is(i)-idiff
        endif
c10     print *, iter, is(j)
 12     continue
c       ibm=ibm*mult
c       index=ishft(ibm,-62)
c       noise
c       is(j)=min0(iq,max0(irand(index)+is(j),1))
 10     continue 
c      if(iter.eq.(iter/1000 )*1000 ) print *, iter,ichange
       if(ichange.eq.0)  goto 11
 9    continue
      print *, 'not converged'
 11   continue
      do 28 i=0,iq
 28     nhist(i)=0
      do 25 i=1,n
        j=is(i)
 25     nhist(j)=nhist(j)+1
c     print *, iter, nhist
      icount=0
      do 27 i=1,iq
        if(nhist(i).gt.0) icount=icount+1
        if(icount.gt.0) number(icount)=number(icount)+nhist(i)
 27   continue
      ict=ict+icount
c     print *, irun,icount,iter
      icount=0
      do 33 i=1,max
 33     icount=icount+number(i)
      derrida=0.0
      fact=1.0/icount**2
      do 34 i=1,max
 34     derrida=derrida+fact*number(i)**2
      do 31 i=1,max
       if(number(i).eq.0) goto 31
       ibin=1+alog(float(number(i)))/0.69315
       ns(ibin)=ns(ibin)+1
 31   continue
c     print *, irun,icount,iter,iq,derrida
      derrisum=derrisum+derrida
 17   continue
      derrida=derrisum/nrun
c     do 32 i=1,31
c32   if(ns(i).gt.0) print *, 2**(i-1), ns(i)
      call flush(6)
 35   print *, idis, ict*1.0/nrun, derrida
      stop
      end
\end{verbatim}

{\bf \large References}
 
ALBERT, R. and Barab\'asi, A.L. (2002), ``Statistical mechanics of
Complex networks'', Rev. Mod. Phys. 74, pp. 47-97.\\

AXELROD, R. (1997), ``The Dissemination of Culture: A Model with Local
Convergence and Global Polarization'', J. Conflict Resolut. 41, pp. 203-226;
{\it The Complexity of Cooperation: Agent-Based Models of Competition
  and Collaboration}, Princeton University Press, Princeton NJ.\\

BARAB\'ASI, A.L. (2002), {\it Linked: The New Science of Networks},
Perseus Books Group, Cambridge MA.\\

BEN-NAIM, E., Krapivsky, P. and Redner, S. (2003), ``Bifurcations and
Patterns in Compromise Processes'', Physica D 183, pp. 190-204.\\

DAVIDSEN, J., Ebel, H. and Bornholdt, S. (2002), ``Emergence of a
small world from local interactions: Modeling acquaintance networks'',
Phys.Rev.Letters 88, pp. 128701-1-128701-4.\\

DEFFUANT, G., Amblard, F., Weisbuch G. and Faure, T. (2002), ``How can
extremism prevail? A study based on the relative agreement interaction
model'', Journal of Artificial Societies and Social Simulation 5 (4), 
paper 1 (http://jasss.soc.surrey.ac.uk/5/4/1.html).\\

DEFFUANT, G., Neau, D. Amblard, F. and Weisbuch, G. (2000), ``Mixing
beliefs among interacting agents. Advances in Complex Systems'',
Adv. Complex Syst. 3, pp. 87-98.\\

DERRIDA, B. and Flyvbjerg, H. (1986), ``Multivalley structure in
Kauffman's model: analogy with spin glasses'', J. Phys. A 19, pp. L1003-L1008.\\

GALAM, S. (1990), ``Social paradoxes of majority rule voting and renormalisation
group'', J. Stat. Phys. 61, pp. 943-951 (1990).\\

GALAM, S. (1997), ``Rational Group Decision Making: a random field
Ising model at T=0'', Physica A 238, pp. 66-80.\\

HEGSELMANN, R. and Krause, U. (2002), ``Opinion Dynamics and Bounded
Confidence Models, Analysis and Simulation'', Journal of Artificial
Societies and Social Simulation 5 (3), paper 2 
(http://jasss.soc.surrey.ac.uk/5/3/2.html).\\

HEGSELMANN, R. and Krause, U. (2004), `` Opinion Dynamics Driven by
Various Ways of Averaging'', Physica A (in press) = 
http://pe.uni-bayreuth.de/?coid=18.\\

HOLME, P. and Kim, B.J. (2002), ``Growing scale-free networks with
tunable clustering'', Phys. Rev. E 65, pp. 026107-1-026107-4.\\

KRAUSE, U. (1997), ``Soziale Dynamiken mit vielen Interakteuren. Eine
Problemskizze''. In Krause, U. and St\"ockler, M. (Eds.), {\it Modellierung und
Simulation von Dynamiken mit vielen interagierenden Akteuren}, pp. 37-51,
Bremen University, Bremen.\\

MOSS DE OLIVEIRA, S., de Oliveira, P.M.C. and Stauffer, D. (1999):
  {\it Evolution, Money, War and Computers}, Teubner, Stuttgart and Leipzig.\\

SCHULZE, C. (2003), ``Advertising in the Sznajd Marketing Model'',
Int. J. Mod. Phys. C 14, pp. 95-98.\\

SCHULZE, C. (2004), ``Advertising, consensus, and ageing in multilayer
Sznajd model'', Int. J. Mod. Phys. C 15, No. 4 (in press) = cond-mat/0312342.\\

SCHWEITZER, F. (2003), {\it Brownian Agents and Active Particles}, Springer, Berlin.\\

STAUFFER, D. (2002), ``Monte Carlo simulations of Sznajd models'',
Journal of Artificial Societies and Social Simulation 5 (1),
paper 4 (http://jasss.soc.surrey.ac.uk/5/1/contents.html).\\

STAUFFER, D. (2003), ``Sociophysics Simulations'', Computing in
  Science and Engineering 5 (3), pp. 71-75; ``How to Convince Others?
  Monte Carlo Simulations of the Sznajd Model'' , in AIP Conference
  Proceedings, {\it The Monte Carlo Method on the Physical Sciences: Celebrating the
  50th Anniversary of the Metropolis Algorithm}, edited by Gubernatis,
  J.E., 690 (1), pp. 147-155 = cond-mat/0307133.\\

STAUFFER, D., Sousa, A.O and Moss de Oliveira, S. (2000),
``Generalization to Square Lattice of Sznajd Sociophysics Model'',
Int. J. Mod. Phys. C  11, pp. 1239-1245.\\

STAUFFER, D. and Meyer-Ortmanns, H. (2004), ``Simulation of Consensus
Model of Deffuant et al on a Barabasi-Albert Network'',
Int. J. Mod. Phys. C 15 (2) (in press) = cond-mat/0308231.\\

SZAB\'O, G., Alava, M. and Kert\'esz, J. (2003), ``Structural
transitions in scale-free networks'', Phys. Rev. E 67, 
pp. 056102-1-056102-1.\\

SZNAJD-WERON, K. and Sznajd, J. (2000), ``Opinion Evolution in Closed
Community'', Int. J. Mod. Phys. C 11, 1157.\\

SZNAJD-WERON, K. and Weron, R.(2003), ``How effective is advertising
in duopoly markets?'', Physica A 324, pp. 437-444.\\

WEIDLICH, W. (2000), {\it Sociodynamics; A Systematic Approach to Mathematical
  Modelling in the Social Sciences}, Harwood Academic Publishers, Amsterdam.\\

WEISBUCH,  G., Deffuant, G.,  Amblard, F. and Nadal, J.-P. (2002),
``Meet, Discuss, and Segregate!'', Complexity 7, pp. 55-63.\\

WEISBUCH, G. (2004), ``Bounded confidence and social networks'',
Eur. Phys. J. B, {\it Special Issue: Application of Complex Networks in
Biological Information and Physical Systems} (in press) = cond-mat/0311279 .

\end{document}